\documentclass[article, nofootinbib,twocolumn]{revtex4-2} 
\usepackage{amsmath}  
\usepackage{amsfonts} 
\usepackage{graphicx} 
\usepackage{bm}        
\usepackage{amssymb}   
\usepackage{float}
\usepackage{physics,mathtools}
\usepackage{slashed}
\usepackage{booktabs}
\usepackage{xcolor,colortbl,tabularx}
\usepackage{epstopdf}
\usepackage{bbold}

\newcommand{\beq}{\begin{equation}}
\newcommand{\eeq}{\end{equation}}
\newcommand{\bea}{\begin{eqnarray}}
\newcommand{\eea}{\end{eqnarray}}

\newcommand{\orcid}[1]{\href{https://orcid.org/#1}{\includegraphics[height=1.9ex,width=1.9ex]{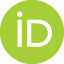}}}

\def\CZ{{\mathcal Z}}
\def\CH{{\mathcal H}}

\begin{document}

\title{Calculating and resumming the classical virial expansion using automated algebra}

\author{Aaron M. Miller\ \orcid{0009-0008-2344-0240}}
\affiliation{Department of Physics and Astronomy, University of North Carolina, Chapel Hill, North Carolina 27599, USA}

\author{Joaqu\'in E. Drut\ \orcid{0000-0002-7412-7165}}
\affiliation{Department of Physics and Astronomy, University of North Carolina, Chapel Hill, North Carolina 27599, USA}

\begin{abstract}
Using schematic model potentials, we calculate exactly the virial coefficients of a classical gas up to sixth 
order and use them to assess the convergence properties of the virial expansion of basic thermodynamic quantities
such as pressure, density, and compressibility. At sufficiently strong couplings, as expected, the virial expansion fails 
to converge. However, at least for the interactions and parameter ranges we explored, we find that Pad\'e-Borel 
resummation methods are extremely effective in improving the convergence of the expansion.
\end{abstract}

\date{\today}

\maketitle

\section{Introduction}

Understanding the finite-temperature thermodynamics of interacting matter represents an important and
challenging problem across many areas of physics and chemistry. Notable applications are
the dynamics of neutron star mergers (where the finite-temperature equations of state of neutron matter
and nuclear matter play a central role), see e.g.~\cite{annurev-nucl-102711-095018, Baiotti:2019sew}, and ultracold atomic gases~\cite{RevModPhys.80.885, RevModPhys.80.1215} (highly malleable systems created 
in many laboratories around the world). While both of those applications involve quantum matter, there are cases
in chemistry and nuclear physics which are better suited for a classical description (usually at high-enough temperature that quantum effects are irrelevant or whenever those can be encoded into effective interactions), see e.g.~\cite{10.1063/5.0113730}. Similarly, classical dynamics simulations of neutron matter at finite temperature have also been of interest~\cite{Horowitz:2011cn}.
This work focuses on such classical descriptions of many-particle systems at finite temperature.

At high temperatures and low densities, the virial expansion (VE) provides a rigorous approach to many-body equilibrium thermodynamics whereby each successive order $N$ adds on the contribution of the $N$-body problem to the 
grand-canonical description. Notably, in recent years the quantum VE has attracted considerable 
attention, in particular in connection with ultracold atomic gases~\cite{Liu2013PR, condmat7010013}, but also as a way to characterize finite-temperature neutron star matter in dilute regimes (see e.g.~\cite{HOROWITZ200655, HOROWITZ2006153, HOROWITZ2006326}). Similarly, as explained in Ref.~\cite{10.1063/5.0113730}, there is also considerable activity in this direction in the area of chemistry, where the last decade has seen renewed interest in virial equations of state.
In all of these cases, novel automated algebra approaches have enabled the calculation of high-order virial coefficients, allowing for the successful application of resummation techniques (see e.g.~\cite{Hou2020PRL, Hou2020PRA}).

In this work, we focus on the application of the VE to a classical gas with a schematic interaction
featuring a purely repulsive two-body force as well as a repulsive force with an attractive pocket at intermediate distances. 
Within the context of that interaction, we explore varying temperatures and coupling strengths in three spatial dimensions
(although, as we explain below, our method is capable of calculating the VE in arbitrary dimensions).
For this purpose, we have developed an automated algebra approach to the calculation of high-order
VE coefficients (based on the seminal work of Ref.~\cite{AaronsHonorsThesis}), which is now available online at~\cite{CVEE}.
For the specific form of the schematic interaction considered here, our results for the coefficients of the VE are exact (up to numerical accuracy limitations) and therefore free of statistical effects (as no stochastic estimators are used in any way).

The remainder of this paper is organized as follows. Section~\ref{Sec:Formalism} presents the formalism
of the VE for a gas of identical particles, first in general form and then specializing to
classical statistics. Section~\ref{Sec:CompMethod} explains the details of our approach to calculating the 
virial coefficients in an automated fashion. In Section~\ref{Sec:Results}, we show the schematic model interaction and 
corresponding results. Finally, in Section~\ref{Sec:Conclusion}, we summarize, conclude, and comment on the
outlook of our work.

\section{Formalism\label{Sec:Formalism}}

The VE organizes the many-body problem into a sum of $N$-body problems,
specifically by Taylor-expanding the grand-canonical partition function $\CZ$ in powers of the fugacity $z$, such that
\beq
\CZ = \sum_{N=0}^{\infty} Q_N z^N,
\eeq
where $Q_N$ is the $N$-particle canonical partition function, $z = e^{\beta \mu}$, $\beta$ is the inverse temperature,
and $\mu$ is the chemical potential. The grand thermodynamic potential $\Omega$ is then given by
\beq
-\beta \Omega = \ln \CZ = Q_1 \sum_{N=1}^{\infty} b_N z^N,
\eeq
where $b_N$ are the virial coefficients
\bea
b_1 &=& 1,\\
b_2 &=& \frac{Q_2}{Q_1} - \frac{Q_1}{2!},\label{Eq:b2}\\
b_3 &=& \frac{Q_3}{Q_1} - b_2 Q_1 - \frac{Q^2_1}{3!},\\
b_4 &=& \frac{Q_4}{Q_1} - \left(b_3 + \frac{b^2_2}{2} \right) Q_1 - b_2\frac{Q^2_1}{2!} - \frac{Q^3_1}{4!}\label{Eq:b4},
\eea
and so on.

To make the connection to the quantum case more explicit, it is worth noting that in that case the expansion coefficients
encode both quantum as well as interaction effects. Indeed, the coefficients of {\it noninteracting}
quantum gases are generally non-vanishing, whereas their classical counterparts are all zero beyond $b_1$.

In the above expressions for $b_N$, the main contribution comes from the term $Q_N/Q_1$; the role of the remaining 
terms is to cancel out contributions from $Q_N$ that scale with super-linear powers of the spatial volume $V$. Once those cancellations are properly accounted for, the final result for $b_N$ is volume-independent. In practice, 
this property implies that one can focus exclusively on those terms in $Q_N$ that are proportional to $V$ 
[since $Q_1$ scales as $V$; see Eqs.~(\ref{Eq:b2})-(\ref{Eq:b4})]. We use this property in our calculations, as further explained below.

For a classical gas of $N$ identical particles in $d$ spatial dimensions, the canonical partition function is 
\beq
Q_N = \frac{1}{N! h^{dN}} \int d^{dN} {\bf p}\int d^{dN} {\bf r}\ e^{-\beta \CH[\{\bf p\}, \{\bf r\}]},
\eeq
where
\beq
\CH[{\{\bf p\}, \{\bf r\}}] = \sum_{i=1}^N \frac{{\bf p}_i^2}{2m} + \sum_{i < j} v_{ij}.
\eeq
Here, ${\bf p}_i$ represents the momentum of the $i$-th particle, ${\bf r}_i$ its position, $m$ its mass (which will be assumed to be the 
same for all particles), and $v_{ij} = v(|{\bf r}_i - {\bf r}_j|)$ is the interaction potential energy that depends on the distance between particle $i$ and particle $j$. We focus
in this work on pairwise interactions, but generalizations to three-body forces and beyond are possible.
In contrast to the quantum case, where momentum and position operators do not commute (and one must resort
to Trotter-Suzuki factorizations; see e.g.~\cite{Hou2019PRA}), here the momenta can be integrated out, which yields
\beq
Q_N = \frac{1}{\lambda_T^{dN} N!}Z_N,
\eeq
where $\lambda_T = \sqrt{2\pi \hbar^2 \beta / m}$ is the thermal wavelength and we define the configuration integral
\beq
Z_N = \int d^{dN} {\bf r}\exp \left({-\beta \sum_{i < j} v_{ij}} \right).
\eeq

Capturing the interaction effects on the grand canonical partition function $\CZ$ through the $b_N$'s amounts to calculating the 
interaction-induced change 
\beq
\Delta Q_N = \frac{1}{\lambda_T^{dN} N!}\Delta Z_N,
\eeq
where
\beq
\Delta Z_N = \int d^{dN} {\bf r}\left[ \exp \left({-\beta \sum_{i < j} v_{ij}} \right) - 1\right].
\eeq

In turn, the above $\Delta Q_N$ determine the change in the virial coefficients $\Delta b_N$,
which enter into the thermodynamics via
\beq
-\beta \Delta \Omega = \ln (\CZ/{\CZ}_0) = Q_1 \sum_{N=1}^{\infty} \Delta b_N z^N,
\eeq
where 
\bea
\Delta b_1 &=& 0,\\
\Delta b_2 &=& \frac{\Delta Q_2}{Q_1},\\
\Delta b_3 &=& \frac{\Delta Q_3}{Q_1} - \Delta b_2 Q_1,
\eea
and so forth, where we have used the fact that $\Delta Q_1 = 0$, since interactions only act among
at least two particles. 

The formalism presented above is the standard one due to Mayer~\cite{MayerMayer} and often found in textbooks 
(see e.g.~\cite{Huang_1987, pathria1972statistical}), albeit not always presented in as much detail as here. The above formulas apply to
any two-body interaction. Below we show how our computational 
method organizes the calculation of $\Delta Z_N$ to access $\Delta b_N$.

\section{Computational method\label{Sec:CompMethod}}

\subsection{Basic considerations}

In order to calculate the central quantities $\Delta Z_N$, we use Mayer's definition of the so-called $f$ function~\cite{MayerMayer}
given by
\beq
\label{Eq:VF}
e^{-\beta v_{ij}} = 1 + f_{ij},
\eeq
such that
\bea
\label{Eq:DeltaZN}
\Delta Z_N = \int d^{dN} {\bf r} \left[ \prod_{i < j} \left(1 + f_{ij}\right) - 1\right].
\eea

The product in this equation has $\binom{N}{2}$ factors and hence $2^{\binom{N}{2}}$ individual terms. 
One of these terms contains no $f$ functions (thus representing a noninteracting contribution) and is equal to unity, 
which will cancel out with the $-1$ term in the square bracket, 
thus leaving $2^{\binom{N}{2}}-1$ total terms in the integrand. Letting $n$ denote the number of $f$ functions 
that appear in a given term, each integrand is a product of the form $f_{i_1 j_1} f_{i_2 j_2} \cdots f_{i_n j_n}$, where 
$1\leq i_k < j_k \leq N$ and $1 \leq n \leq \binom{N}{2}$.

Making the reasonable assumption that the interaction is translation-invariant, one may always
factor out the center-of-mass motion, which upon integration shows that $\Delta Z_N$ scales at least as $V$.
For the VE coefficients to remain finite, any terms scaling as a power of $V$ higher than linear must ultimately be cancelled out in 
the final expression for $\Delta b_N$ (otherwise the VE coefficients would be infinite in the thermodynamic limit) and can 
therefore be discarded right away. The computational job thus starts with selecting the terms in $\Delta Z_N$ that scale only 
linearly with $V$. To that end, note that any product of $f$ functions that does not contain all available indices 
$1, \dots, N$ will yield scaling with $V$ beyond linear and can therefore be discarded. 
Even if all the indices do appear in a given term, one must ensure that they do not form disjoint subsets, i.e. it must not be possible 
to factor the product of $f$'s into two (or more) sub-factors containing disjoint sets of indices. We comment more systematically on 
these properties below, after introducing a graph-based notation.

Following the definition used in~\cite{Huang_1987}, we establish a bijective correspondence between the integrals in 
Eq.~(\ref{Eq:DeltaZN}) and undirected $N$-particle graphs. Let the nodes of an $N$-particle graph be labeled $1,2,\dots,N$. 
Given an arbitrary integral term, for each factor $f_{i_k j_k}$ appearing in the integrand (which has the form 
$f_{i_1 j_1} f_{i_2 j_2} \cdots f_{i_n j_n}$), connect an undirected edge between nodes $i_k$ and $j_k$. We then say 
that the graph represents the integral 
\beq
\int d^{d} {\bf r}_1 d^{d} {\bf r}_2 \cdots d^{d} {\bf r}_n f_{i_1 j_1} f_{i_2 j_2} \cdots f_{i_n j_n}.
\eeq

\begin{figure}[h]
	\centering
	\includegraphics[width=0.5\linewidth]{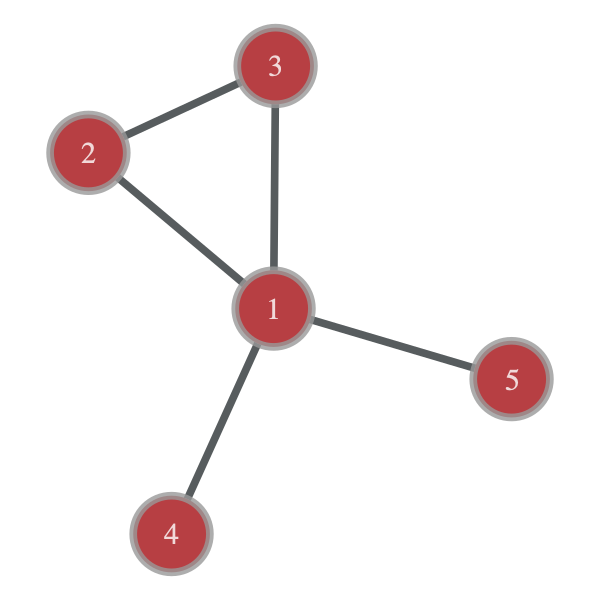}   
	\includegraphics[width=0.5\linewidth]{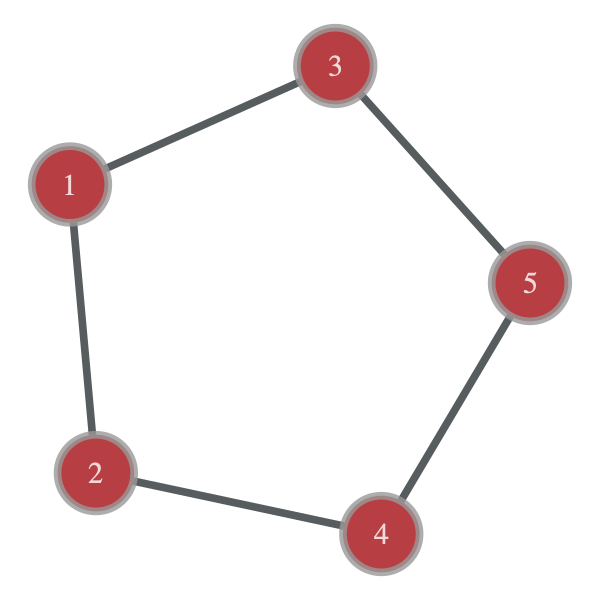}    
	\caption{
		\label{fig:Graphs}
		Two contributions at order $N=5$, $n=5$ that scale as $V$ and hence contribute to $\Delta b_5$.
		}
\end{figure}

As examples, we show in Fig.~\ref{fig:Graphs} two contributions at order $N=5$, $n=5$ that scale as $V$ (and therefore
contribute to the final result for $\Delta b_5$).  Explicitly, they represent the integrals
\beq
\int d^{d} {\bf r}_1 d^{d} {\bf r}_2  d^{d} {\bf r}_3 d^{d} {\bf r}_4 d^{d} {\bf r}_5
f_{1 2} f_{1 3} f_{1 4} f_{1 5} f_{2 3},
\eeq
and
\beq
\int d^{d} {\bf r}_1 d^{d} {\bf r}_2  d^{d} {\bf r}_3 d^{d} {\bf r}_4 d^{d} {\bf r}_5
f_{1 2} f_{1 3} f_{2 4} f_{4 5} f_{3 5},
\eeq
respectively.

\begin{figure}[h]
	\centering
	\includegraphics[width=0.5\linewidth]{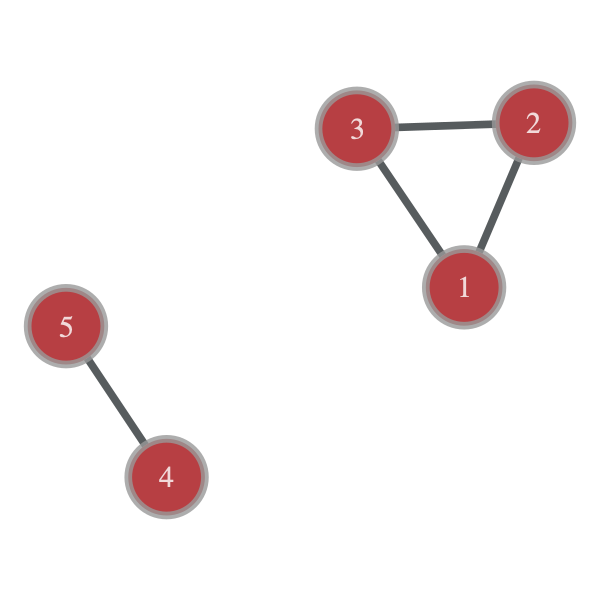}   
	\caption{
		\label{fig:GraphDisconnected}
		A contribution at order $N=5$, $n=4$ that scales as $V^2$, which
		cancels out in the final expression for $\Delta b_5$.
		}
\end{figure}

As an example of a term that does not contribute to the final answer for $\Delta b_5$, we show in 
Fig.~\ref{fig:GraphDisconnected} a disconnected graph that represents the factorable integral
\beq
\label{Eq:disjoint}
\int d^{d} {\bf r}_1 d^{d} {\bf r}_2  d^{d} {\bf r}_3 f_{1 2} f_{2 3} f_{1 3}
\int d^{d} {\bf r}_4 d^{d} {\bf r}_5 f_{4 5}.
\eeq

%

\subsection{Systematic identification of contributing graphs}

Since we are interested in the terms in $\Delta Z_N$ that scale as $V$, we start by considering all simple $N$-particle graphs with $n$ edges, $1 \le n \le \binom{N}{2}$, corresponding to the $2^{\binom{N}{2}}-1$ terms in Eq.~(\ref{Eq:DeltaZN}). The restriction to simple graphs excludes self-loops and multiple edges, for each factor $f_{ij}$ must have distinct indices and can appear at most once in a given integrand.

As explained above, any term will scale at least linearly in $V$ due to translational invariance, and any term that admits a disjoint factorization [e.g. Eq.~(\ref{Eq:disjoint})] will scale faster than $V$ and therefore not contribute to $\Delta b_N$. In the graphical representation, such factorability corresponds to a disconnected graph. Hence, any disconnected graph may be ignored. Since a graph must have at least $N-1$ nodes to be connected, this means that we can immediately discard all graphs with $n<N-1$ in our calculations. Next, assuming that all particles interact via the same pairwise potential, the result of each integral is invariant to permutation of the indices. Expressed graphically, this means that any two isomorphic graphs will yield the same numerical value upon computation. Graph isomorphism is an equivalence relation on a set of graphs, so we can partition the set of graphs with $n\ge N-1$ into isomorphism equivalence classes such that the integral terms in each are numerically equivalent. 

At this point, we discard all graphs in equivalence classes representing disconnected diagrams (Discarding all graphs with 
$n<N-1$ was necessary but not sufficient for removing all disconnected graphs). We are thus left with a set of simply connected 
graphs partitioned into isomorphism classes. We can represent this collection with a ``multiset" 
\beq
    \label{redg}
    \mathcal{G}^*  = \{g_1G_1,g_2G_2,\dots\},
\eeq
where the elements $G_i$ represent unique isomorphism class representatives and the repetition numbers $g_i$ represent the 
size of the corresponding isomorphism class. After numerically evaluating the $G_i$, we can form the vectors 
$\boldsymbol{g} = [g_1\ g_2\ \dots]^T$ and $\boldsymbol{G} = [G_1\ G_2\ \dots]^T$ and calculate Eq.~(\ref{Eq:DeltaZN}) via
\beq
    \label{lc}
    \Delta Z_N = \boldsymbol{g}^T\boldsymbol{G}.
\eeq
Our implementation contains the combinatorial data of $\mathcal{G}^*$ (and hence can compute the VE coefficients) beyond the 
sixth order presented here. The data through order six were obtained by brute force isomorphism testing using the \texttt{graph-
tools} library, which implements the VF2 algorithm of Cordella et al. (see \cite{1323804}). The data for higher orders were obtained 
using the repository from Ref.\cite{CombinatorialData} and the fact that
\beq
g_i = \frac{N!}{|\mathrm{Aut}(G_i)|},
\eeq
where $|\mathrm{Aut}(G_i)|$ is the cardinality of the \textit{auto}morphism class of an $N$-particle graph $G_i$.

\subsection{Integral evaluation}

Evaluating Eq.~(\ref{lc}) requires us to calculate each of the $dN$-dimensional integrals appearing in $\boldsymbol{G}$. A 
completely general treatment would involve constructing the functional form of each integrand and proceeding with a broadly 
applicable numerical integration method such as Newton-Cotes or Monte Carlo. We choose instead to base our scheme on the 
multivariate Gaussian identity
\beq
\label{mgauss}
  \int\exp(-\frac{1}{2}\boldsymbol{x}^T M \boldsymbol{x})d^m\boldsymbol{x}=\sqrt{\frac{(2\pi)^m}{\det M}},
\eeq
where $M\in R^{m\times m}$ is a symmetric positive definite matrix and $x\in R^m$. While this identity corresponds to an exact 
integration only for potentials $v_{ij}$ that yield an integrand of this form when transformed through Eq.~(\ref{Eq:VF}), the class of 
potentials for which this method is exact can emulate both a purely repulsive interaction and an interaction with an attractive 
pocket, as we describe below. 

Additionally, this method is computationally efficient. The integrands never have to be constructed explicitly, and the results may be 
obtained by computing an $N\times N$ determinant, even though the integration space is $dN$-dimensional. Indeed, computing 
the VE coefficients to ninth order, which requires evaluating $\mathcal{O}(10^5)$ unique integrals, can be accomplished in 
seconds. Lower orders can be achieved almost instantaneously, allowing one to plot smooth curves describing the evolution of the 
coefficients.

Another advantage of Eq.~(\ref{mgauss}) is that it permits the use of non-integer dimensions. In our case, the dimension enters the 
result via $m=dN$ and the power to which we raise the eigenvalues of one block of the quadratic form represented by $A$. Thus, 
one could simply regard $d$ as a parameter of the investigation, such that examining how the VE coefficients change with 
dimension would amount to the exponentiation of scalars, sidestepping the already small cost of the determinant computations.

As a final note, our current implementation offers to approximate a general integrand with one of the form in Eq.~(\ref{mgauss}). 
We have done this to provide some degree of generality to others who may use our code, but the general case is not the focus of 
this paper. We restrict our attention moving forward to those potentials for which the results are exact.



\section{Model and Results\label{Sec:Results}}

In this section we present our results for the pressure and density equations of state, as well as the isothermal compressibility.
The approach detailed in the previous sections applies to an arbitrary two-body potential $v_{ij}$ where, in general, the integrals that result will not have exact analytic forms (i.e. one will not be able to use the simple result valid for Gaussians, mentioned above), such that a stochastic evaluation is needed, with the concomitant statistical uncertainties. To avoid such uncertainties, 
in this work we use a class of schematic model potentials for which an exact evaluation is possible. Specifically, we define our 
two-body interaction potential $v_{ij}$ to be such that
\beq
\label{Eq:FIJ}
f_{ij} = A e^{-b_1({\bf r}_i - {\bf r}_j)^2} - (1 + A) e^{-b_2({\bf r}_i - {\bf r}_j)^2},
\eeq
where $A, b_1, b_2$ are constants; we will refer to this assumption as a Gaussian model. In other words, 
rather than fixing the shape of $v_{ij}$ and setting the inverse temperature $\beta$, and extracting $f_{ij}$ from them
via Eq.~(\ref{Eq:VF}), in this work we test our calculations by fixing the constants $A, b_1, b_2$ above, thus letting
the interaction to be dictated by
\beq
v_{ij} = -\frac{1}{\beta}\ln \left(1 + f_{ij}\right).
\eeq
In a realistic application, one would instead take $v_{ij}$ as an input and determine the temperature
dependence of $A, b_1, b_2$ by optimization.

\begin{figure}[h]
	\centering
	\includegraphics[width=\linewidth]{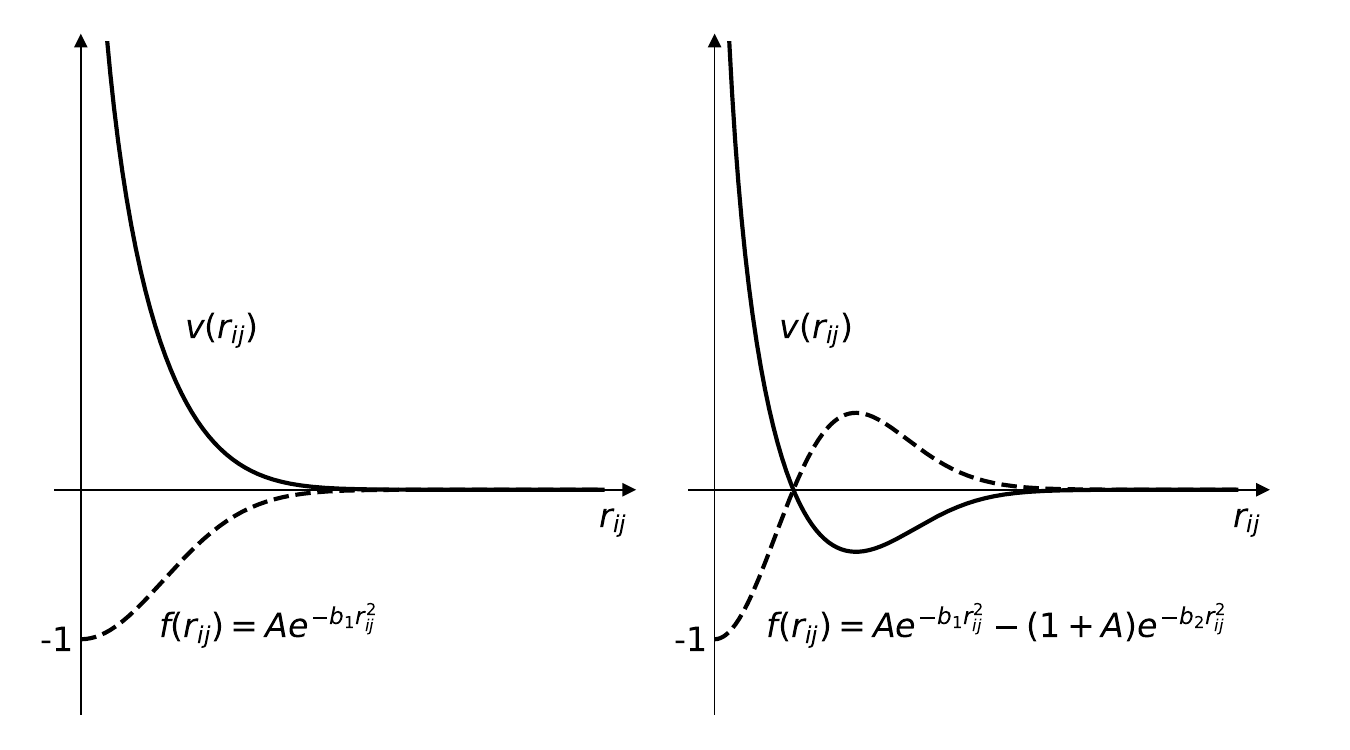}   
	\caption{
		\label{fig:Potentials}
		Left panel: Solid line shows the potential $v_{ij}$ as a function of $r_{ij} = |{\bf r_i} - {\bf r_j}|$, 
		resulting from the Gaussian model 
		of Eq.~(\ref{Eq:FIJ}) setting $A=-1$, $b_1>0$, shown with a dashed line.
		Right panel: More general case corresponding to arbitrary $A > 0$ and $b_2 > b_1 > 0$ 
		(or $A < -1$ and $b_1 > b_2 > 0$).
		}
\end{figure}

With $f_{ij}$ given by Eq.~(\ref{Eq:FIJ}), one may easily represent physically interesting situations such as 
a repulsive interaction (setting $A = 0$ or $A = -1$), as well as a repulsive two-body potential 
with an attractive pocket (for $A>0$ or $A<-1$). In this work we explore both of those situations, shown schematically in Fig.~\ref{fig:Potentials}.
%


\subsection{Purely repulsive interaction}\label{purerep}

\begin{figure}[h]
	\centering
	\includegraphics[width=\linewidth]{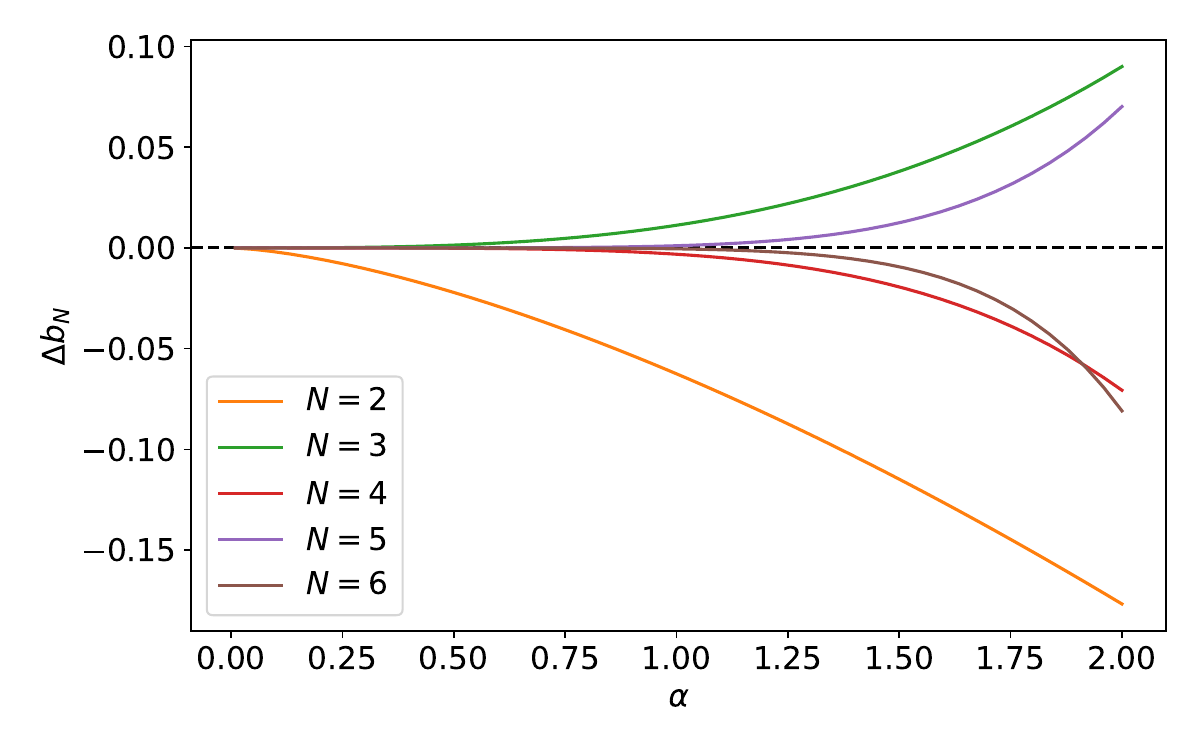}   
	\caption{
		\label{fig:VECoefficients}
		Virial coefficients $\Delta b_N$ for $N=2$,$3$,$4$,$5$,$6$ for the repulsive model ($A=-1$) as a function of 
		the dimensionless coupling $\alpha = T/b_1$.
		}
\end{figure}

In Fig.~\ref{fig:VECoefficients} we show our results for the virial coefficients $\Delta b_N$ of the purely repulsive Gaussian 
model (i.e. $A=-1$; see left panel of Fig.~\ref{fig:Potentials}), as a function of the dimensionless coupling $\alpha = T/b_1$.
As $\alpha$ grows, so do the interaction effects on the virial coefficients, as expected. In particular, we see how at large 
enough $\alpha$, high-order coefficients tend to become larger than their lower-order counterparts; this type of behavior was 
found as well in the quantum case in Refs.~\cite{Hou2020PRL, Hou2020PRA} and it reflects the breakdown of the convergence 
properties of the series.
With the virial coefficients in hand, it is straightforward to evaluate the partial sums up to the available order
for the pressure, density, and isothermal compressibility. They are given by
\bea
\frac{P}{P_0} &=& 1+\sum_{N=2}^\infty\Delta b_N z^{N-1}, \label{Eq:pp0}\\
\frac{n}{n_0} &=& 1+\sum_{N=2}^\infty N\Delta b_Nz^{N-1}\label{Eq:nn0},
\eea
and
\beq
 \frac{\kappa}{\kappa_0} = \frac{1+\displaystyle\sum_{N=2}^{\infty}N^2\Delta b_Nz^{N-1}}{\left(1+\displaystyle\sum_{N=2}^{\infty}N\Delta b_Nz^{N-1}\right)^2},
\eeq
where in all cases the subscript $0$ indicates the noninteracting case and we have used the thermodynamic identity
\beq
\label{Eq:kk0}
\kappa = \frac{\beta}{n^2}\left . \frac{\partial n}{\partial (\beta\mu)}\right |_T.
\eeq
It is straightforward to evaluate Eqs.~(\ref{Eq:pp0}) and (\ref{Eq:nn0}) as partial sums, but doing so for Eq.~(\ref{Eq:kk0}) requires a bit more care. As written, Eq.~(\ref{Eq:kk0}) will include partial contributions to higher orders, making it unclear what such an expression represents (e.g. Evaluating Eq.~(\ref{Eq:kk0}) with coefficients up to $\Delta b_3$ will not result in a quadratic plot, but will instead include the contributions of $\Delta b_2$ and $\Delta b_3$ to terms that are cubic and quartic in the fugacity). Hence, to keep Eq.~(\ref{Eq:kk0}) on the same footing as Eq.~(\ref{Eq:pp0}) and Eq.~(\ref{Eq:nn0}), we rewrite it as a single power series 
\beq
\label{Eq:kk0new}
\frac{\kappa}{\kappa_0} = \sum_{N=1}^{\infty}c_Nz^{N-1},
\eeq
where $c_1 = 1$ and 
\bea
c_2 &=& 0,\\
c_3 &=& 3\Delta b_3-4\Delta b_2^2,\\
c_4 &=& 8\Delta b_4-24\Delta b_2\Delta b_3+16\Delta b_2^3,
\eea
and so on. Eq.~(\ref{Eq:kk0new}) truncated at $c_N$ then represents the largest partial sum of $\kappa/\kappa_0$ exactly computable using VE coefficients up to order $N$. [The analogous interpretation holds for Eqs.~(\ref{Eq:pp0}) and (\ref{Eq:nn0}).]

\begin{figure}[h]
	\centering
	\includegraphics[width=\linewidth]{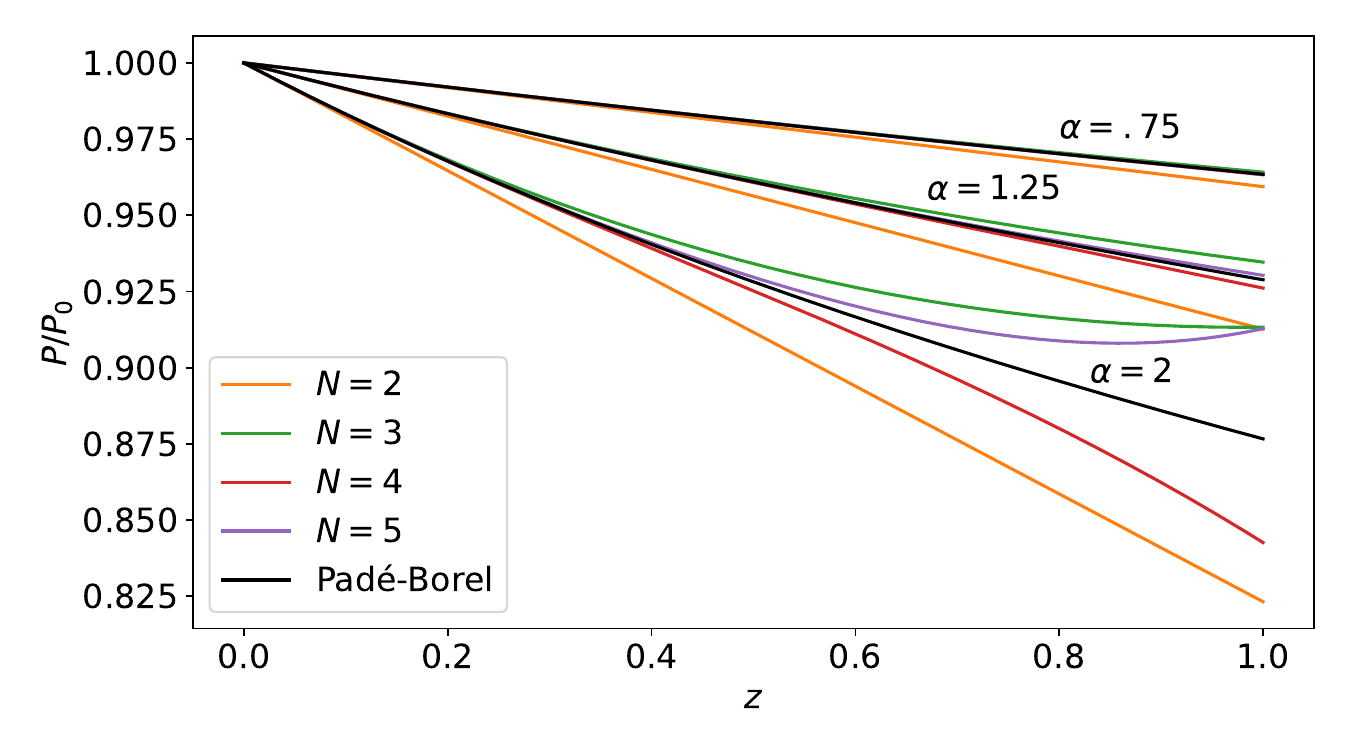}   
	\caption{
		\label{fig:Pressure}
		Pressure $P$ for the purely repulsive model, in units of its noninteracting counterpart $P_0$, as a function of the fugacity $z$ for three representative values of the dimensionless coupling $\alpha = T/b_1$. The colored lines show the highest value of the virial coefficient included, following the same convention as in Fig.~\ref{fig:VECoefficients}. The black line shows the result of a Pad\'e-Borel resummation, described below.
		}
\end{figure}

\begin{figure}[h]
	\centering
	\includegraphics[width=\linewidth]{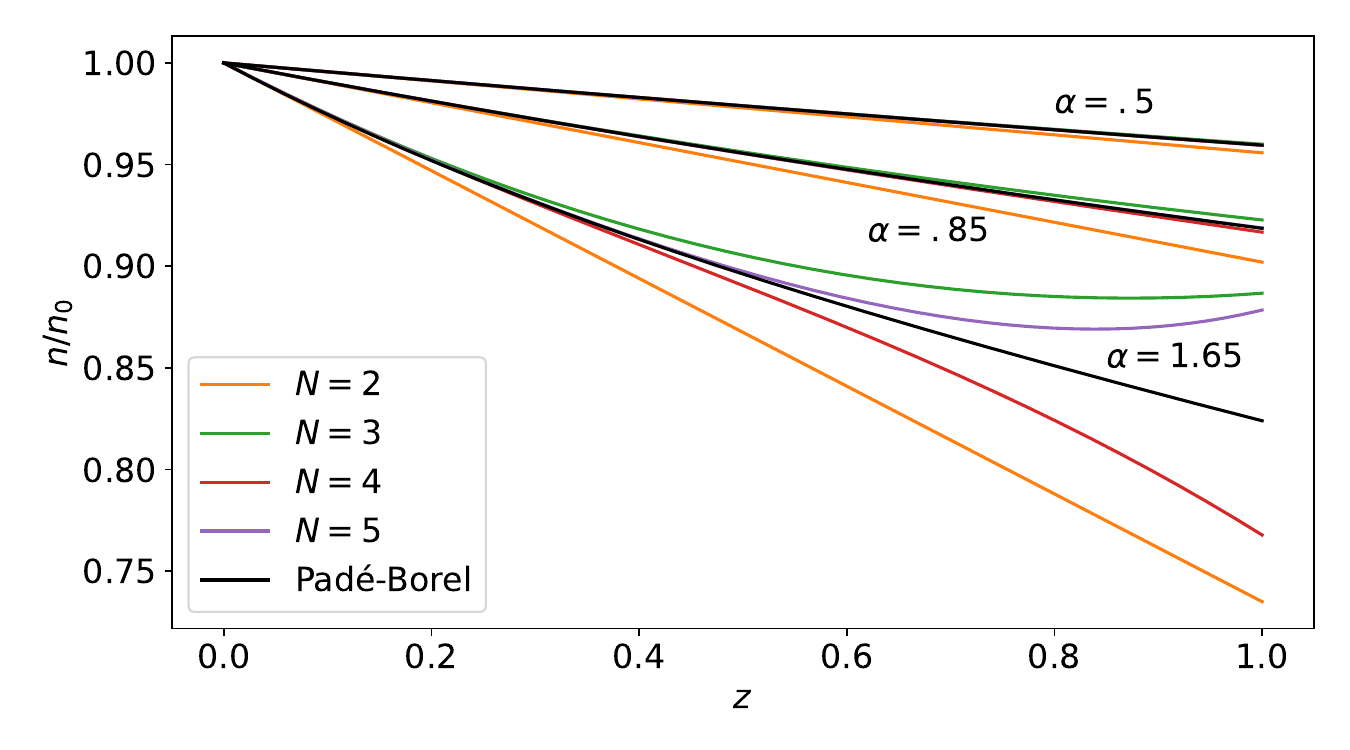}   
	\caption{
		\label{fig:Density}
		Density $n$ for the purely repulsive model, in units of its noninteracting counterpart $n_0$, as a function of the fugacity 
		$z$ for three representative values of the dimensionless coupling $\alpha = T/b_1$. The colored lines show 
		the highest value of the virial coefficient included, following the same convention as in Fig.~\ref{fig:VECoefficients}. 
		The black line shows the result of a Pad\'e-Borel resummation, described below.
		}
\end{figure}

\begin{figure}[h]
	\centering
	\includegraphics[width=\linewidth]{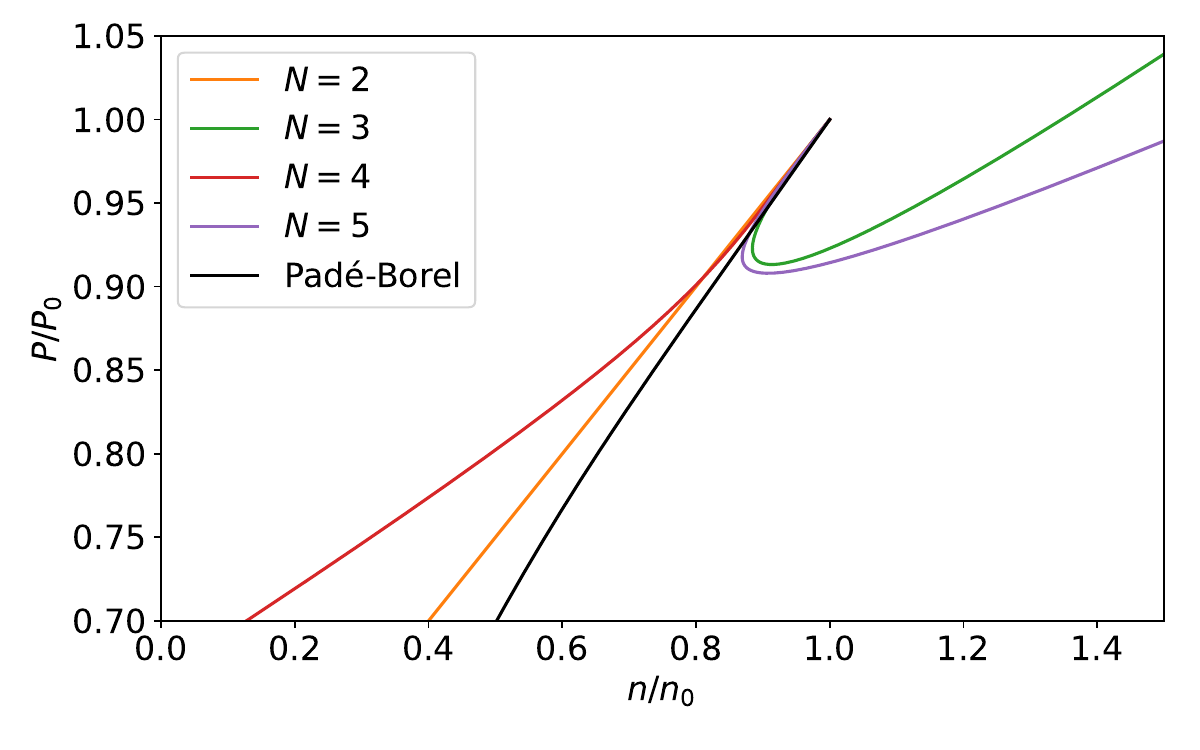}   
	\caption{
		\label{fig:PressureDensity}
		Pressure-density equation of state for the purely repulsive model at $\alpha = T/b_1 = 1.25$. The colored lines show 
		the highest value of the virial coefficient included, following the same convention as in Fig.~\ref{fig:VECoefficients}. 
		The black line shows the result of a Pad\'e-Borel resummation, described below.
		}
\end{figure}

\begin{figure}[h]
	\centering
	\includegraphics[width=\linewidth]{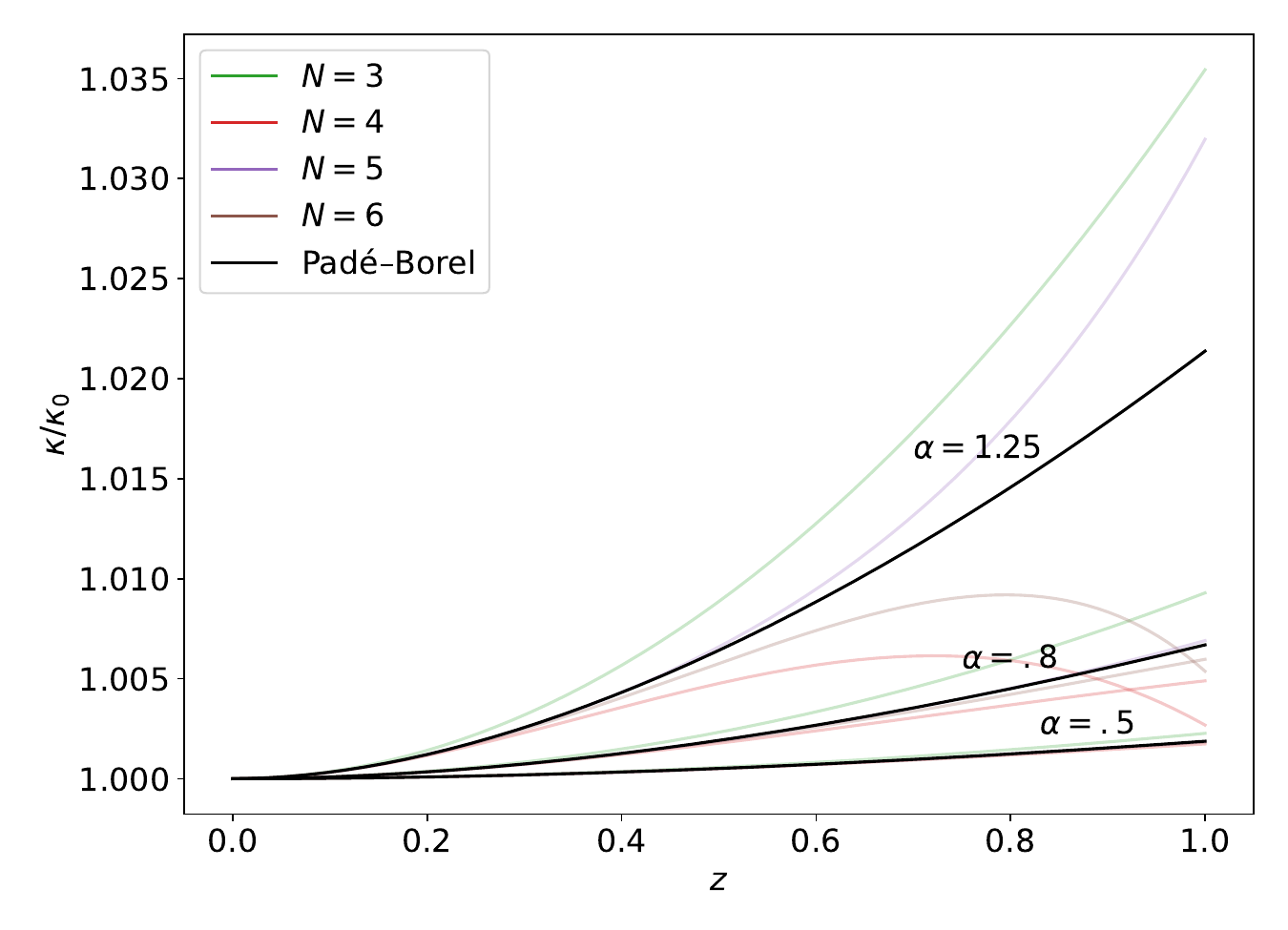}   
	\caption{
		\label{fig:Compressibility}
		Compressibility $\kappa$ for the repulsive model, in units of its noninteracting counterpart $\kappa_0$, as a function of the fugacity $z$. The black line shows the result of a Pad\'e-Borel resummation.
		}
\end{figure}

Figures~\ref{fig:Pressure} through~\ref{fig:Compressibility} show the above quantities for representative values of $\alpha$.
It is often useful to display the pressure-density equation of state, which amounts to a parametric plot that combines
the information in Figs.~\ref{fig:Pressure} and \ref{fig:Density}. We show such a plot in Fig.~\ref{fig:PressureDensity}.
%

\subsubsection{Pad\'e-Borel resummation of the VE}

For strong enough interactions (in the sense of sufficiently large $\alpha$), the partial sums of the VE show 
clear signs of convergence failure. To address this issue, we resort to resummation methods, specifically Pad\'e-Borel 
resummation. In this approach, one replaces a given power series (in our case the VE)
\beq
g(z) = \sum_{n=0}^{\infty} a_n z^n,
\eeq
with its Borel transform, namely
\beq
Bg(z) = \sum_{n=0}^{\infty} \frac{a_n}{n!} z^n,
\eeq
whose convergence properties can be expected to be more favorable than those of the partial sums of the original function $g(z)$. 
Using the highest available partial sum for $Bg(z)$, a Pad\'e approximant is used as an ansatz to fit the resulting function. These 
approximants take the rational form $P(z)/Q(z)$, where $P$ and $Q$ are polynomials. Once a proper fit is obtained (in particular 
one that does not display poles for real values of $z$, which would be unphysical), the resummed function is obtained (in fact, 
defined within the context of the resummation method) via
\beq
\label{Eq:inv}
g(z) = \int_0^{\infty} e^{-t} Bg(zt) dt,
\eeq
which is evaluated numerically.

In each plot featuring a Pad\'e-Borel resummation, the Pad\'e approximant is fitted to the Borel transform of the highest partial sum 
displayed in the figure. Also, for all Pad\'e approximants, $P(z)$ is linear and $Q(z)$ is quadratic. We experimented with different 
polynomials orders for $P$ and $Q$ but found that this combination behaved most reliably when performing the inverse transform 
of Eq.~(\ref{Eq:inv}). The linear-quadratic approximant can replicate the Borel transforms of the pressure and density partial sums 
well, but the approximation quality lessens for the compressibility due to its increased curvature.

Figures~\ref{fig:Pressure} through~\ref{fig:Compressibility} display the result of carrying out a Pad\'e-Borel resummation on the 
series for the pressure, density, and compressibility. Our results show that this resummation approach vastly improves the
convergence properties of the VE (at least for the quantities and parameter ranges studied).

\subsection{Repulsive interaction with attractive pocket}

\begin{figure}[h]
	\centering
	\includegraphics[width=\linewidth]{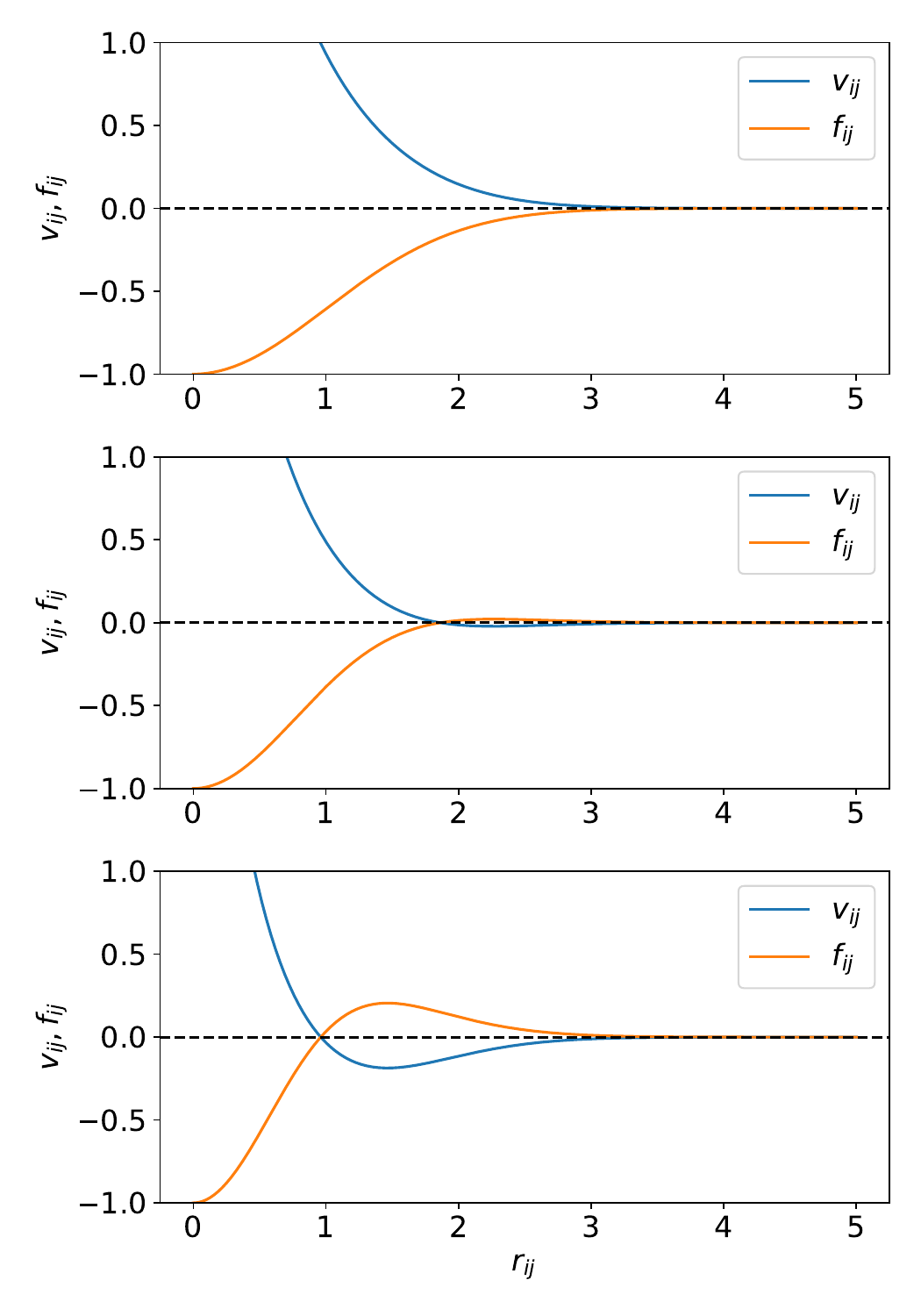}   
	\caption{
		\label{fig:Potentials2}
		Potential $v_{ij}$ as a function of $r_{ij} = |{\bf r_i} - {\bf r_j}|$ (the horizontal
		axis is the same for all three plots), 
		resulting from the Gaussian model of Eq.~(\ref{Eq:FIJ}) setting $A=1$
		and varying the value of $\gamma = b_2/b_1$. From top to bottom,
		the plots show the cases of $\gamma = 1, 1.4,$	 and $2.5$, respectively.
		}
\end{figure}

Encouraged by the results obtained for the purely repulsive interaction, we analyze here the more interesting case
of an interaction that is repulsive at short distances but includes the more realistic feature of having an attractive pocket.
We obtain the latter from Eq.~(\ref{Eq:FIJ}) by setting $A=1$ (an arbitrary illustrative choice) and varying values of 
$\gamma = b_2/b_1$. At $\gamma = 1$, the contribution from $A$ disappears in Eq.~(\ref{Eq:FIJ}) and one recovers the repulsive
case considered in the previous section. As $\gamma$ is increased beyond 1, an attractive pocket develops 
in the interaction (at a rate governed by the value of $A$), as shown qualitatively in Fig.~\ref{fig:Potentials} and
Fig.~\ref{fig:Potentials2}.

\begin{figure}[h]
	\centering
	\includegraphics[width=\linewidth]{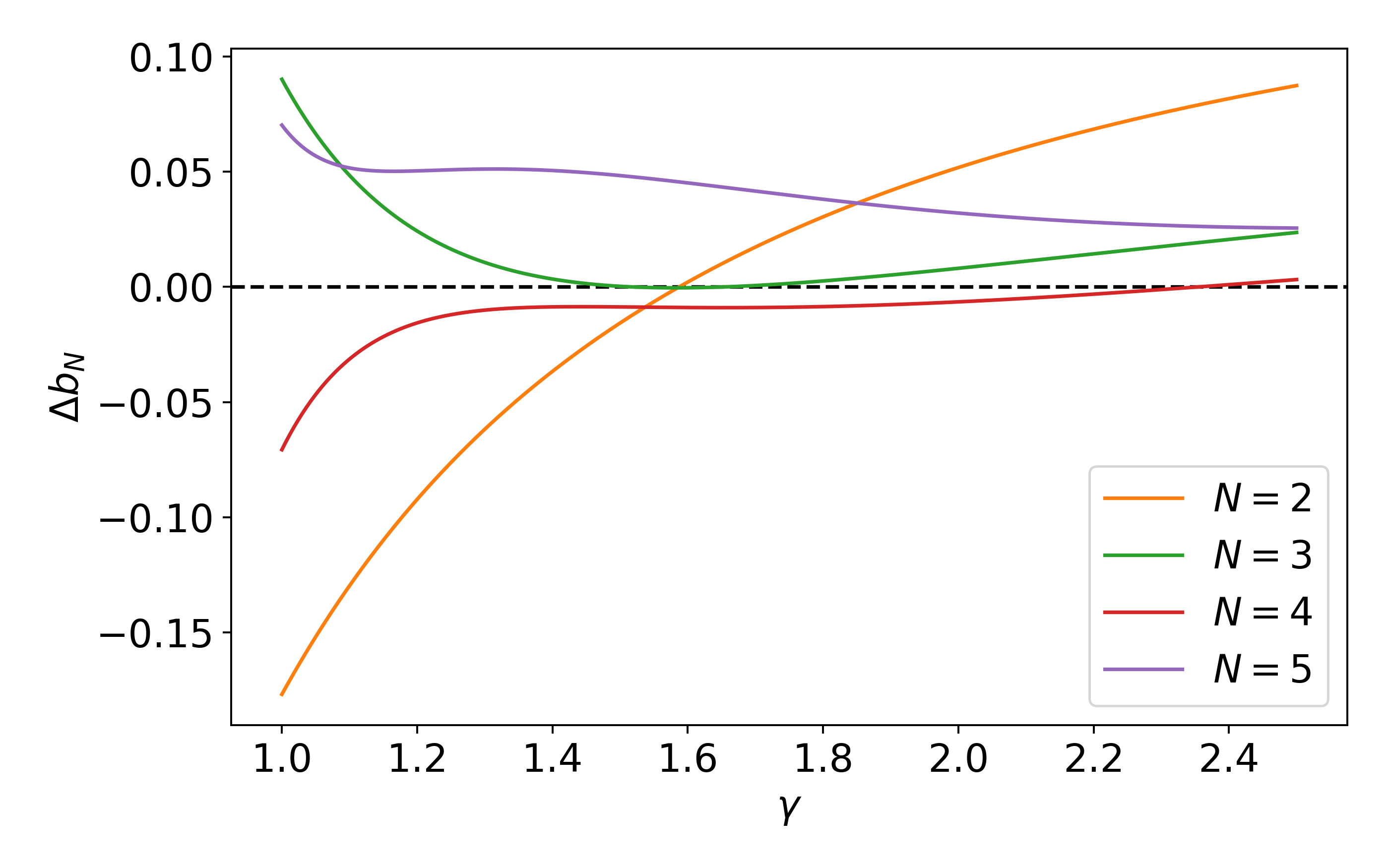}   
	\caption{
		\label{fig:VECoefficientsAttractive}
		Virial coefficients $\Delta b_N$ for $N=2$, $3$, $4$, $5$ for the repulsive model with an attractive pocket 
		($A=1$, $\alpha = 2$) as a function of the dimensionless coupling $\gamma = b_2 / b_1$. At $\gamma=1$,
		the model becomes the purely repulsive limit of the previous section. As $\gamma$ is increased beyond 1,
		the attractive pocket develops, as shown in Fig.~\ref{fig:Potentials2}.
		}
\end{figure}

In Fig.~\ref{fig:VECoefficientsAttractive}, we show the results of our calculations for the virial coefficients
as a function of $\gamma$ at $\alpha = T/b_1 = 2$ (the latter being the strongest coupling considered in the 
previous section).

\begin{figure}[h]
	\centering
	\includegraphics[width=\linewidth]{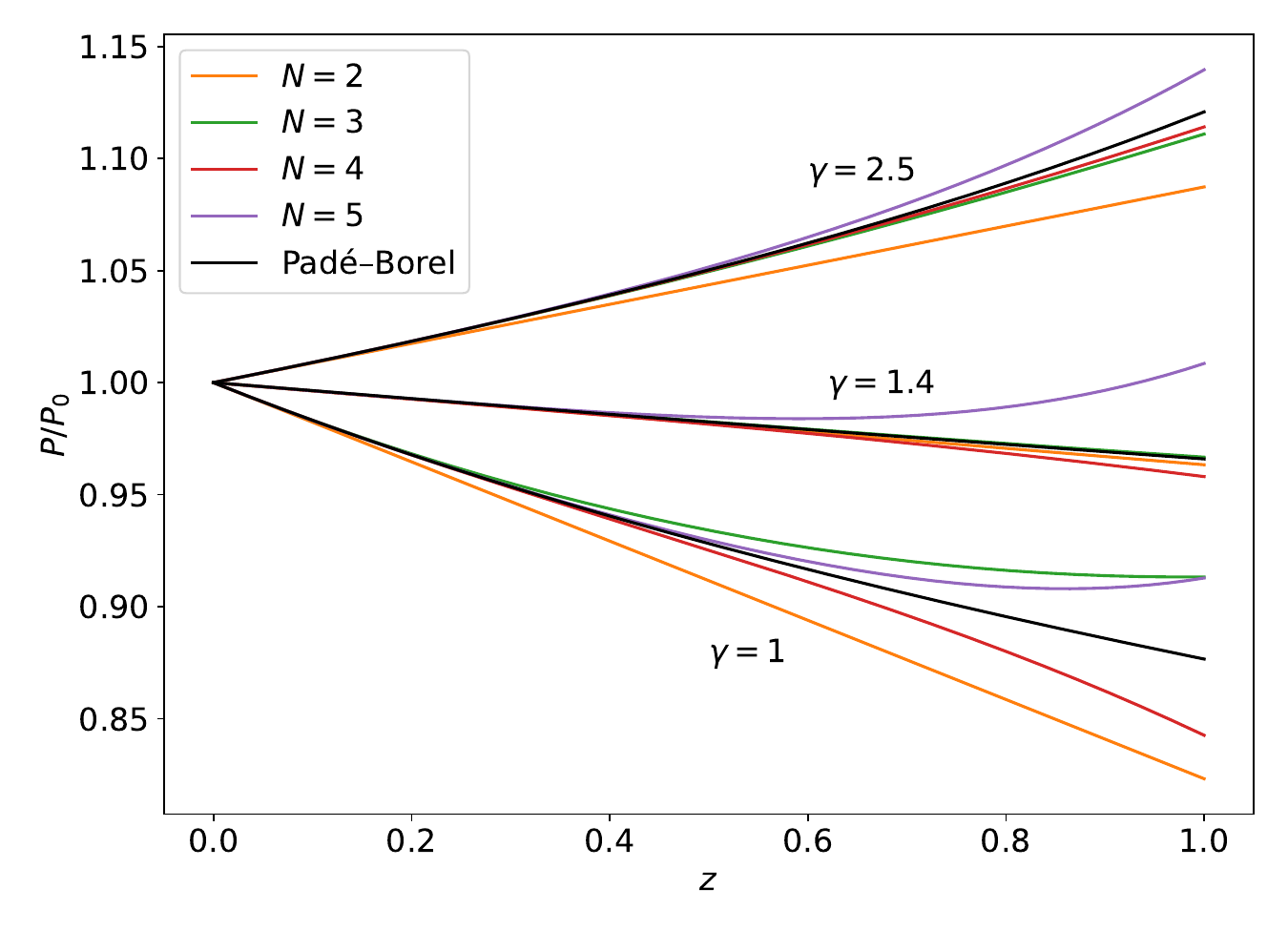}   
	\caption{
		\label{fig:Pressure2}
		Pressure $P$ for the repulsive model with attractive pocket, in units of its noninteracting counterpart $P_0$, as a function of the fugacity $z$, for three different values of the interaction parameter $\gamma$. Note that $\gamma = 1$ corresponds to the purely repulsive case studied in the previous section. The black line shows the result of a Pad\'e-Borel resummation.
		}
\end{figure}

\begin{figure}[h]
	\centering
	\includegraphics[width=\linewidth]{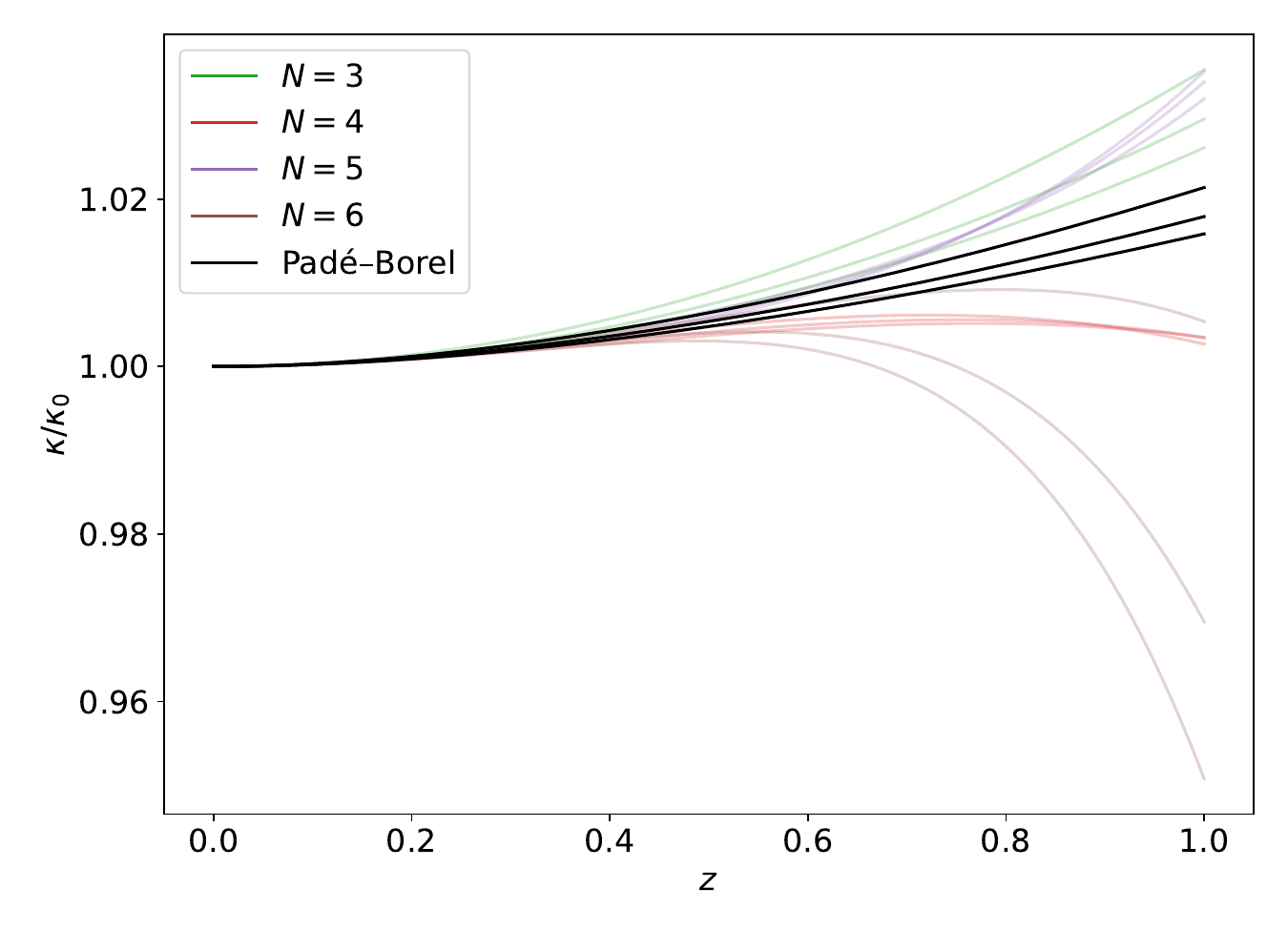}   
	\caption{
		\label{fig:Compressibility2}
		Compressibility $\kappa$ for the repulsive model with attractive pocket, in units of its noninteracting counterpart $\kappa_0$, as a function of the fugacity $z$. The black lines show the result of a Pad\'e-Borel resummation, from top to bottom
for $\gamma = 1, 1.03, 1.05$, respectively.
		}
\end{figure}

Following closely the discussion of the previous section, we show in Figs.~\ref{fig:Pressure2} and \ref{fig:Compressibility2} the pressure and compressibility, respectively, as functions of $z$, at fixed $\alpha$ and varying $\gamma \geq 1$, for the model with an attractive pocket. In each figure, the fixed value of $\alpha$ used is the strongest coupling considered in the corresponding purely repulsive case (see Figs.~\ref{fig:Pressure} and \ref{fig:Compressibility}).
Once again, our results show that this resummation approach vastly improves the
convergence properties of the VE (at least for the quantities and parameter ranges studied). 

\section{Conclusion and outlook\label{Sec:Conclusion}}

In this work we have explored, for a schematic model interaction encoding two different physical situations, 
the convergence properties of the VE of a classical gas. To that end, we have implemented an automated algebra 
approach to the calculation of high-order VE coefficients $\Delta b_N$. Using those, we calculated the pressure and 
density equations of state, as well as the isothermal compressibility.

As one of our main conclusions, we have found that resummation techniques such as Pad\'e-Borel can vastly 
extend the applicability of the VE, at least for the class of models and parameter ranges we studied.
Although we present this optimistic view, our results should be taken with the proverbial grain of salt, as the analytic
properties of the VE are not well known for the specific family of models we considered.

Another main result of this work is the creation of an automated algebra package, which can be found online as the 
computational virial expansion engine (CVE$^2$); see Ref.~\cite{AaronsHonorsThesis} and Ref.~\cite{CVEE} for continued 
developments and releases. To the best of our knowledge, this 
is the first project addressing this problem by implementing an approach full based on automated algebra 
(without numerical integration), as presented here. Although we have only used
CVE$^2$ to calculate up to $b_6$ in this work, the code is prepared to go beyond $b_7$ in its present form.

The most straightforward generalizations of our analysis, which will shed further light on the possibilities of the method,
its implementation via CVE$^2$, and the properties of the VE, include extensions to multispecies systems (here we 
focused on identical particles of a single type), and within the latter the possibility of mass imbalance and spin polarization. 
Another aspect worth exploring is the dependence on spatial dimension, which is straightforward in our approach since 
the dimension enters analytically as a variable; in other words, within CVE$^2$, we can study classical gases not only
in three spatial dimensions (as done here) and lower integer dimensions, but also in fractional dimensions, which
may be of interest from the mathematical physics perspective. 

Finally, it is worth pointing out that we explored here a schematic interaction where the Mayer factor $f$ was modeled
as a single Gaussian function or a sum of two Gaussian functions. In future generalizations of this study, a higher number of 
Gaussians could be used to study, for instance, more realistic interactions such as screened Coulomb potentials. We leave such 
investigations to future work.
\\

\acknowledgments

We would like to thank Y. Hou and G. Rogelberg for discussions during the very early stages of this work.
This material is based upon work supported by the National Science Foundation under Grant No. PHY2013078.


\bibliographystyle{apsrev4-2}
\bibliography{CVEERefs}


\end{document}